\input{aipcheck}

\documentclass[
    ,final            
  ]
  {aipproc}

\layoutstyle{8x11single}

\begin{document}

\title{Dirac neutrinos, dark matter stability and flavour predictions from Lepton Quarticity}

\classification{11.10.-z, 11.15.-q, 11.15.Ex, 11.30.-j, 11.30.Fs, 12.15.Ff, 12.60.-i, 12.90.+b, 13.15.+g, 14.60.Pq, 14.60.St, 14.80.Cp }
\keywords      {Dirac neutrinos, Dark Matter, Flavour predictions, Quarticity}

\author{Salvador Centelles Chuli\'a}{
  address={AHEP Group, Institut de F\'{i}sica Corpuscular --
  C.S.I.C./Universitat de Val\`{e}ncia, Parc Cient\'ific de Paterna.\\
 C/ Catedr\'atico Jos\'e Beltr\'an, 2 E-46980 Paterna (Valencia) - SPAIN}
}



\begin{abstract}
We propose to relate dark matter stability to the possible Dirac nature of neutrinos.
The idea is illustrated in a simple scheme where small Dirac neutrino masses arise
from a type-I seesaw mechanism as a result of a $Z_4$ discrete lepton number symmetry.
The latter implies the existence of a viable WIMP dark matter candidate, whose
stability arises from the same symmetry which ensures the Diracness of neutrinos.
The symmetry groups $\Delta_{27}$ and $A_4$ are then used to extract a rich 
variety of flavour predictions.
\end{abstract}

\maketitle


\section{Introduction}

Amongst the major shortcomings of the Standard Model are the neutrino mass and
the dark matter stability problem. Underpinning the origin of neutrino mass and elucidating the
nature of dark matter would constitute a gigantic step forward in particle physics. Here we
focus on the possibility that the \textbf{neutrino nature and dark matter stability problem may be closely
interconnected}. Concerning neutrinos a major unknown is whether they are
their own anti - particles, an issue which has remained an open challenge ever since Ettore
Majorana had his pioneering idea on the quantum mechanics of spin \cite{Majorana:1937vz}. On the other hand, since many
years, physicists have pondered about what is the dark matter made of, and what makes it
stable, a property usually assumed in an ad-hoc fashion. Indeed, although the existence
of non-baryonic dark matter is well established by using cosmological and astrophysical
probes, its nature has otherwise remained elusive. 

Recently it has been argued in \cite{Hirsch:2017col} that the Dirac or Majorana nature of neutrinos is intimately connected 
to the breaking pattern of $U(1)_L$ Lepton number symmetry, which is accidentally conserved in the Standard Model, to its residual conserved $Z_n$ subgroups.
%
%
Interestingly enough, neutrinos can only be Majorana for specific $Z_n$ groups and even within these specific $Z_n$ groups, only for special choices of the charges, while they will be Dirac in any other case. 
Here we focus on the $ U(1)_L \rightarrow {Z}_4$ scenario, which we call \textbf{Quarticity}.
We show that \textbf{a WIMP dark matter candidate can naturally emerge, stabilized by Quarticity,
the same symmetry associated to the Diracness of neutrinos}. Moreover, the smallness of neutrino masses can be understood through the Dirac analogue of type-I seesaw. We also show that many different symmetry groups 
can be used to give the flavour structure of neutrino mixings and therefore \textbf{a rich variety of models arise}, showing different features and predictions but all maintaining the principal result, \textbf{the connection between Dark Matter and the Dirac nature of neutrinos}.

\section{The model}

The general structure of the model is given in \cite{Chulia:2016ngi}. Concerning the particle content of the model, we need to introduce some new fields in addition to those in the Standard Model. In the fermionic sector we introduce \textbf{3 generations 
of right handed neutrinos}, $\nu_{i,R}$, to obtain neutrino masses, as well as \textbf{3 more generations of heavy neutral leptons} with the two chiralities, wich we will call $N_{i, L}$ and $N_{i,R}$,
in order to have a seesaw mechanism.
We also have to introduce a new scalar, $\chi$, which has a \textbf{non - zero vev}. In addition there are two other scalars carrying non-trivial $Z_4$ charges, a real scalar $\eta$, and a complex scalar $\zeta$, which is the \textbf{DM candidate}. All of them are gauge singlets. The set of transformation rules under $Z_4$ for the fields in the model is: 
\pagebreak

  \begin{table}[ht]
\centering
\begin{tabular}{c c || c  c}
    \hline \hline
  \textbf{Fields}            \hspace{5mm}        &  ${Z}_4 $ 		\hspace{5mm}            & 
  \textbf{Fields}            \hspace{5mm}        &  $ Z_4 $	\hspace{5mm}                        \\
  \hline \hline
  \textbf{Fermions}            \hspace{5mm}        &                 \hspace{5mm}            & 
              \hspace{5mm}        &                \hspace{5mm}                        \\
  \hline 
  $\bar{L}_{i,L}$  \hspace{5mm}        &  $\mathbf{z}^3$      \hspace{5mm}               &  
  $\nu_{i,R}$      \hspace{5mm}        &  $\mathbf{z}$        \hspace{5mm}               \\
  $ l_{i,R} $      \hspace{5mm}        &  $\mathbf{z}$        \hspace{5mm}                & 
  $\bar{N}_{i,L}$  \hspace{5mm}        &  $\mathbf{z}^3$      \hspace{5mm}                     \\
  $N_{i,R}$        \hspace{5mm}        &  $\mathbf{z}$        \hspace{5mm}               &                                      
                   \hspace{5mm}        &                      \hspace{5mm}        \\
\hline
  \textbf{Scalars (vev $\neq $0)}            \hspace{5mm}        &                \hspace{5mm}            & 
             \hspace{5mm}        &                \hspace{5mm}                        \\

  $\Phi$         \hspace{5mm}        &  $\mathbf{1}$        \hspace{5mm}                  &
  $\chi$         \hspace{5mm}        &  $\mathbf{1}$        \hspace{5mm}                                \\
\hline
  \textbf{Scalars (vev=0)}            \hspace{5mm}        &               \hspace{5mm}            & 
   	    \hspace{5mm}        &                 \hspace{5mm}                        \\

  $\zeta$          \hspace{5mm}        &  $\mathbf{z}$        \hspace{5mm}                 &          
  $\eta$           \hspace{5mm}        &  $\mathbf{z}^2$      \hspace{5mm}                                \\
    \hline \hline
  \end{tabular}
\caption{Charge assignments for leptons, quarks,  scalars 
  ($\Phi_i^u$, $\Phi_i^d$ and $\chi_i$) as well as ``dark matter sector'' 
  ($\zeta$ and $\eta$). Here $\mathbf{z}$ is the fourth root
  of unity, i.e. $\mathbf{z}^4 = 1$.  }
 \label{tab1} 
\end{table}

The only requirement for the flavour structure of the model is that \textbf{the flavour symmetry group $G$ forbids the tree level neutrino mass}, this is, $\bar{L} \phi^c \nu_R$. Then \textbf{neutrino masses are generated via a type-I seesaw}.

\subsection{Neutrino Diracness}

\textbf{The symmetry rules under ${Z}_4$ ensures that neutrinos are Dirac particles by forbidding all the possible Majorana terms at all orders. }
   Note that the scalars which have a non-zero vev, $\Phi$ and $\chi$, transform trivially under ${Z}_4$, thus implying that \textbf{the Quarticity symmetry ${Z}_4$
   is not spontaneously broken}. Moreover
   
   \begin{equation}
(\Phi^\dagger \Phi)^n \chi^m \sim 1
   \end{equation}

   \noindent This implies that if all the Majorana mass terms are forbidden at tree level, then they are forbidden at all orders. This is indeed the case, because \textbf{all neutral fermionic field transform as $z$ under ${Z}_4$}.
   Therefore, all the Majorana masses at tree level are forbidden. For example
   
   \begin{equation}
    \bar{\nu}_R^c \nu_R \sim z^2
   \end{equation}
   
   \noindent The higher order mass terms -via non-zero vevs of the scalars $\Phi$ and $\chi$- also transform non-trivially under the quarticity symmetry

   \begin{equation}
    \bar{\nu}_R^c \nu_R (\Phi^\dagger \Phi)^n \chi^m \sim z^2
   \end{equation}
  
  Therefore, these terms are forbidden  and \textbf{neutrinos remain Dirac particles at all orders, while the symmetry ${Z}_4$ is not broken after spontaneous symmetry breaking}.

\subsection{A stable Dark Matter candidate}

To ensure that neutrinos remain Dirac particles, the ${Z}_4$ quarticity symmetry should remain exact, so that no scalar carrying a ${Z}_4$ charge should
acquire any vev. Thus \textbf{both $\zeta$ and $\eta$ which carry ${Z}_4$ charge can potentially be stable as a
result of the unbroken quarticity symmetry}. However, owing to the ${Z}_4$ charge assignments, the $\eta$
field has cubic couplings to both scalars and fermions. These
couplings lead to the decay of $\eta$ and thus, despite carrying a ${Z}_4$ charge, \boldmath $\eta$ \unboldmath  \textbf{is not stable}. On the other hand, in the lagrangian there is no term of the form $\zeta \rho_i \rho_j$, where $\rho$ is a generic scalar, nor of the form $\zeta \psi_i \psi_j$, 
 where $\psi$ is a generic fermion. This implies that \boldmath $\zeta$ \unboldmath \textbf{is an stable particle, thus it is a viable dark matter candidate}. \vspace{3mm}
 
\begin{figure}[!h]
 \centering
  \includegraphics[height=4cm]{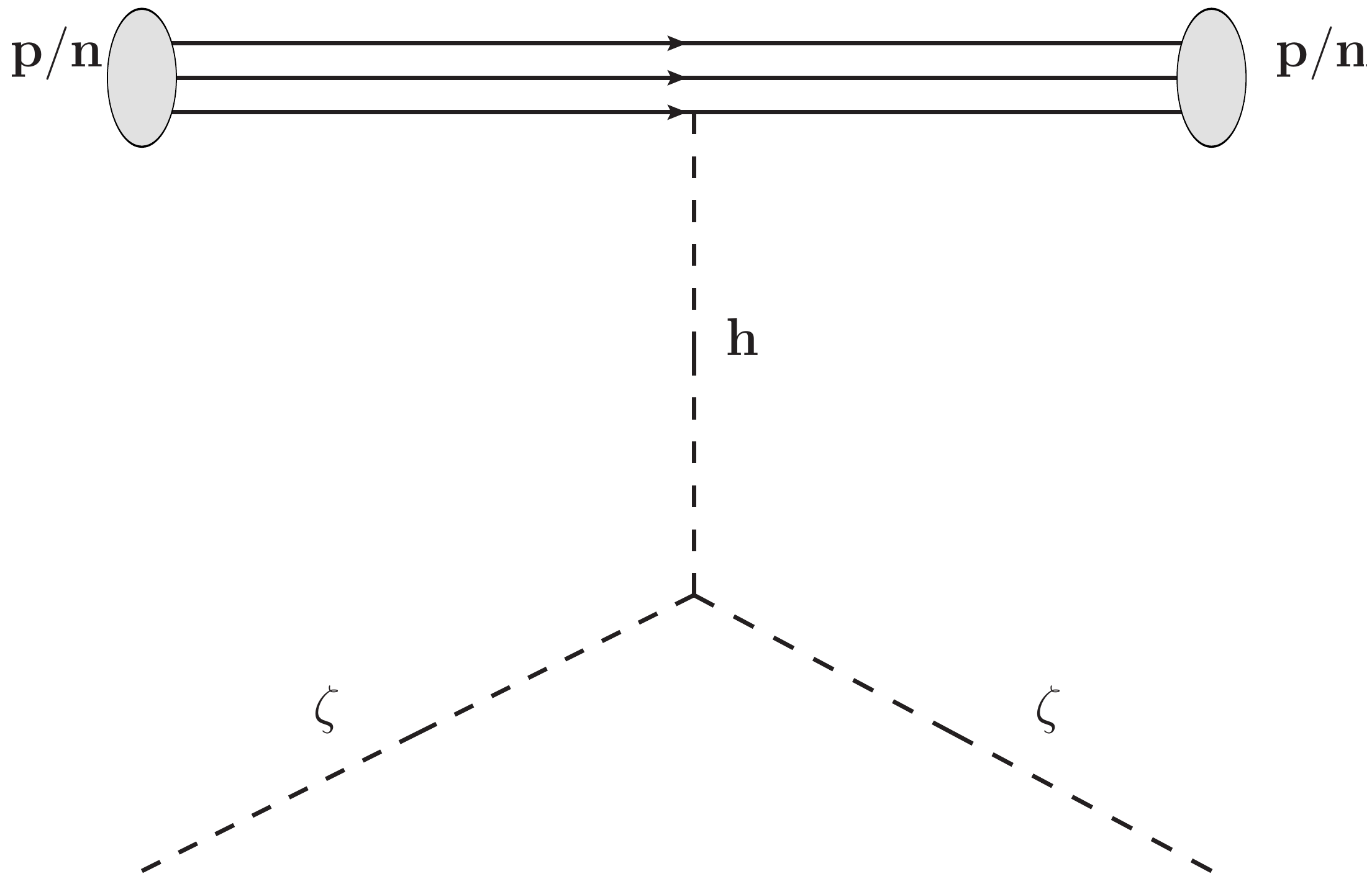} \hspace{10mm}
  \includegraphics[height=4cm]{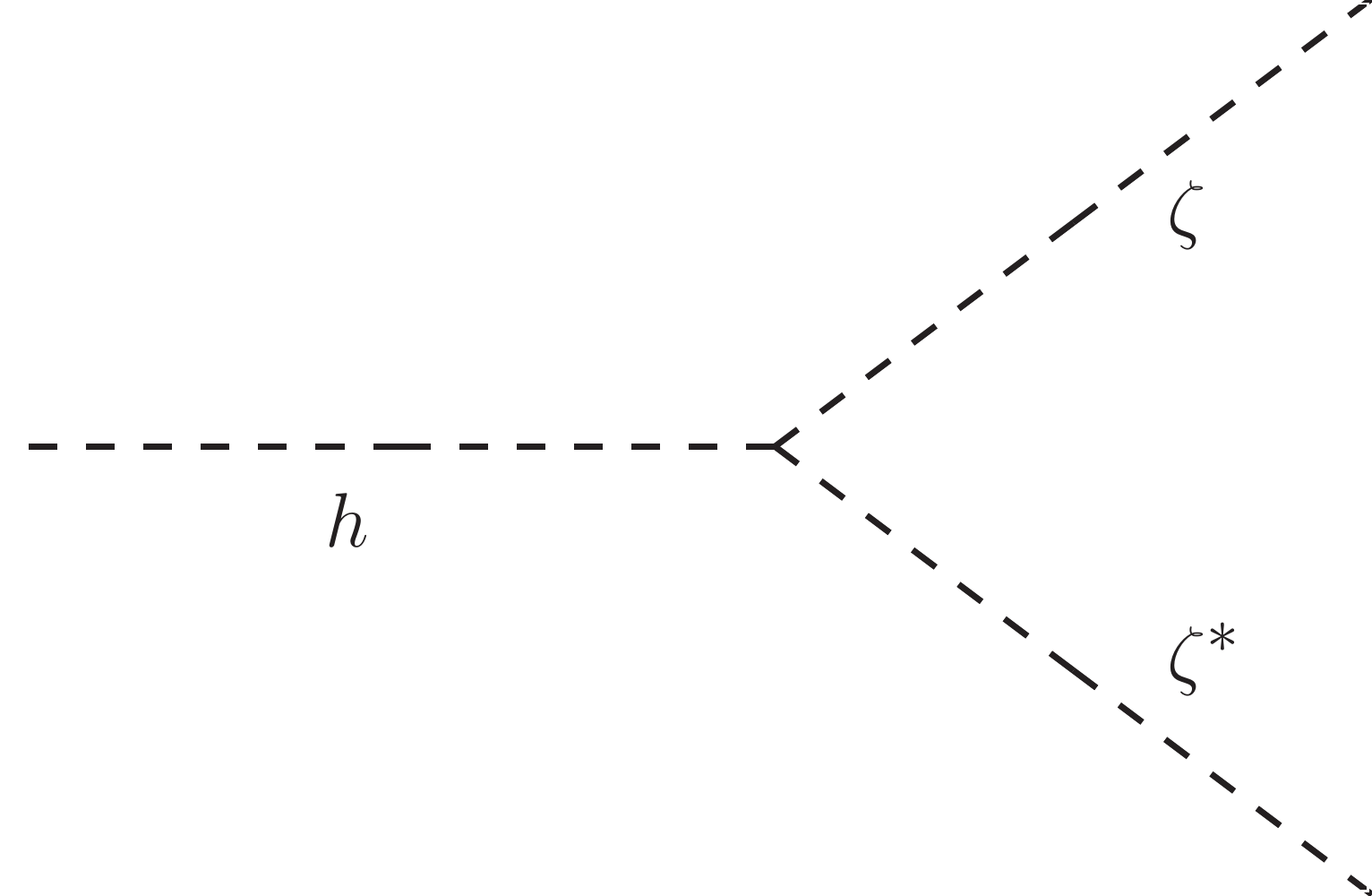}
  \caption{Left panel: Leading order Feynman diagram of the nuclear recoil of the dark matter candidate via a exchange of the Higgs boson. Right panel: Tree level Feynman diagram of the invisible decay of the Higgs boson to a pair of DM particles.}
    \label{fig:experimental}
  \end{figure}

\noindent The detection could be direct, via \textbf{nuclear recoil}. The WIMP particle will also contribute to the \textbf{invisible decay of the Higgs}.
Both features are shown in the following diagrams \ref{fig:experimental}

\begin{figure}[!h]
 \centering
\includegraphics[height=4cm]{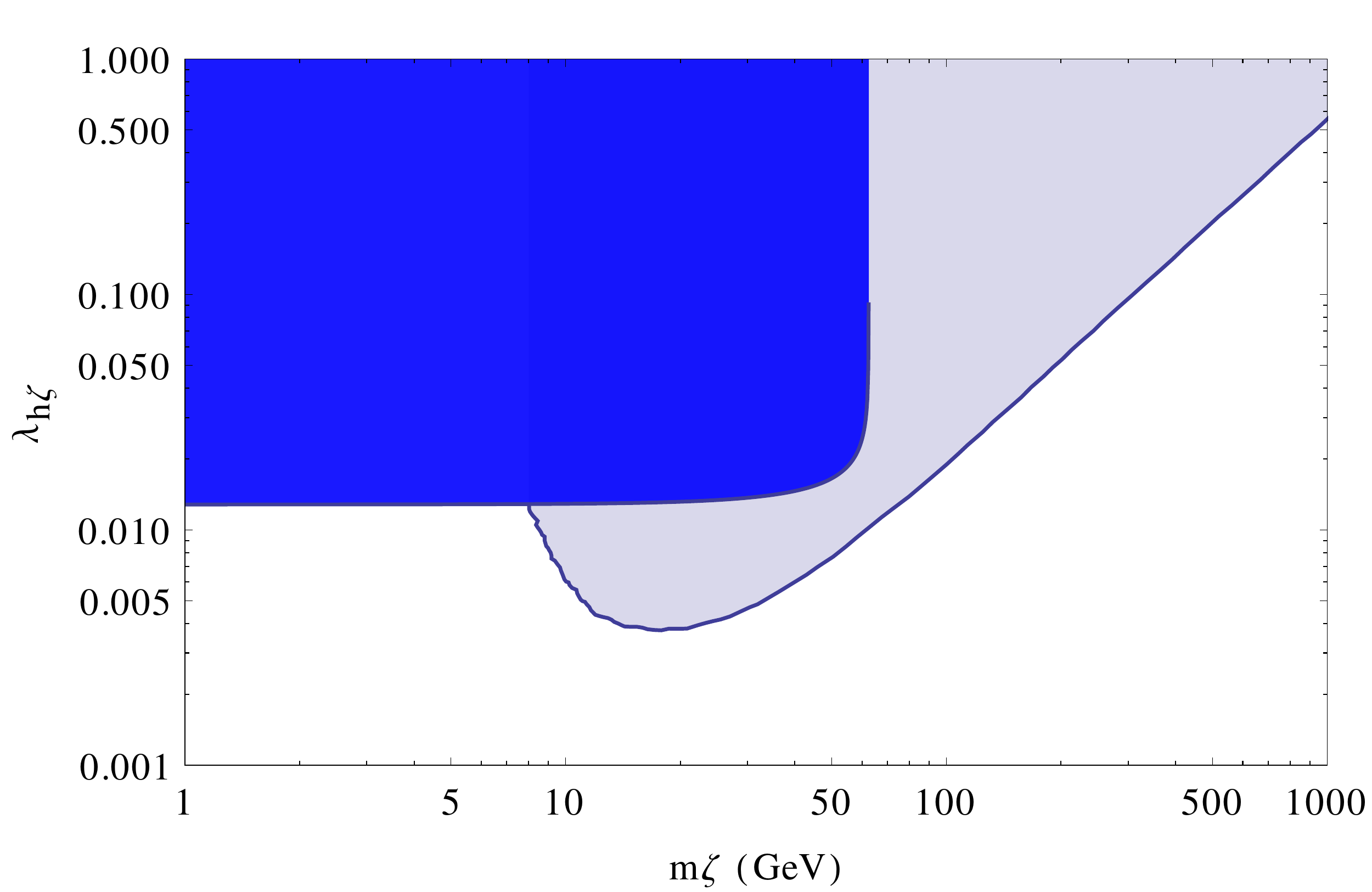}
  \caption{Parameter space excluded by LHC invisible decay data and LUX direct searches in the  $\lambda_{h\zeta}$ vs $m_\zeta$ plane.}
    \label{fig:experimental2}
  \end{figure}

 \noindent \textbf{The constraints from LHC for invisible decay and the for direct searches coming from LUX experiment rules out 
 part of the parameter space regarding the higgs-dark matter quartic coupling $\lambda_{h\zeta}$ and mass of dark matter $m_\zeta$}, which can be seen in the figure \ref{fig:experimental2}.  \vspace{3mm}
 
\section {Flavour predictions}

The model includes, apart from the Standard Model gauge group, a global symmetry $Z_4 \otimes G$, where $Z_4$ is the Quarticity symmetry already discussed while $G$ is the flavour symmetry. It's role is to forbid the tree level coupling between 
left and right handed neutrinos and this can be trivially done using $Z_2$, as done in \cite{Chulia:2016ngi}. However, one can use non-abelian discrete symmetries to extract very interesting flavour predictions
 
\subsection{$\Delta_{27}$ as flavour symmetry}

This model was discussed in \cite{Chulia:2016giq}. \textbf {When we take \boldmath $\Delta(27)$ \unboldmath as the flavour symmetry}, $G$, an interesting structure appears in the neutral fields mass matrix, which can then be diagonalized
  and the masses and mixing parameters can be extracted. The parameters in the model have enough freedom to \textbf{fit all the mixing angles and neutrino masses constraints}. The charged assignments are shown in Table \ref{tab2}.
  
\begin{table}[h]
\centering
\begin{tabular}{c c c || c c c}
  \hline \hline
  Fields           \hspace{1cm}        & $\Delta(27)$          \hspace{1cm}        &  $Z_4$                    \hspace{1cm}      & 
  Fields           \hspace{1cm}        & $\Delta(27)$          \hspace{1cm}        &  $Z_4$                                     \\
  \hline \hline
  $\bar{L}_e$      \hspace{1cm}        & $\mathbf{1}$          \hspace{1cm}        &  $\mathbf{z}^3$      \hspace{1cm}      &  
  $\nu_{e,R}$      \hspace{1cm}        & $\mathbf{1}$          \hspace{1cm}        &  $\mathbf{z}$                          \\
  $\bar{L}_\mu$    \hspace{1cm}        & $\mathbf{1''}$        \hspace{1cm}        &  $\mathbf{z}^3$      \hspace{1cm}      &
  $\nu_{\mu, R}$   \hspace{1cm}        & $\mathbf{1'}$         \hspace{1cm}        &  $\mathbf{z}$                          \\
  $\bar{L}_\tau$   \hspace{1cm}        & $\mathbf{1'}$         \hspace{1cm}        &  $\mathbf{z}^3$      \hspace{1cm}      &   
  $ \nu_{\tau,R}$  \hspace{1cm}        & $\mathbf{1''}$        \hspace{1cm}        &  $\mathbf{z}$                          \\
  $ l_{i,R} $      \hspace{1cm}        & $\mathbf{3}$          \hspace{1cm}        &  $\mathbf{z}$        \hspace{1cm}      & 
  $\bar{N}_{i,L}$  \hspace{1cm}        & $\mathbf{3}$          \hspace{1cm}        &  $\mathbf{z}^3$                        \\
  $N_{i,R}$        \hspace{1cm}        & $\mathbf{3'}$         \hspace{1cm}        &  $\mathbf{z}$         \hspace{1cm}     &                                      
                   \hspace{1cm}        &                       \hspace{1cm}        &                                             \\
  \hline
  $\Phi_i$         \hspace{1cm}        & $\mathbf{3'}$         \hspace{1cm}        &  $\mathbf{1}$             \hspace{1cm}      &
  $\chi_i$         \hspace{1cm}        & $\mathbf{3'}$         \hspace{1cm}        &  $\mathbf{1}$                              \\
  $\zeta$          \hspace{1cm}        & $\mathbf{1}$          \hspace{1cm}        &  $\mathbf{z}$        \hspace{1cm}      &               
  $\eta$           \hspace{1cm}        & $\mathbf{1}$          \hspace{1cm}        &  $\mathbf{z}^2$                        \\
    \hline
  \end{tabular}
\caption{ The $\Delta (27)$ and $Z_4$ charge assignments for leptons,
  the Higgs scalars ($\Phi_i$, $\chi_i$) and the dark matter sector
  scalars ($\zeta$ and $\eta$). Here $\mathbf{z}$ is the fourth root
  of unity, i.e. $\mathbf{z}^4 = 1$.}
  \label{tab2}
\end{table}
  
  The interesting scenario arises when we consider the following alignment for the vevs of the scalars
  
  \begin{eqnarray}
   \langle \Phi_i \rangle = v \hspace{1mm}(1,1,1)\\
   \langle \chi_i \rangle = u \hspace{1mm}(1, 1+\epsilon, 1+\alpha)
  \end{eqnarray}

\begin{figure}[!h]
 \centering
 \includegraphics[height=5cm]{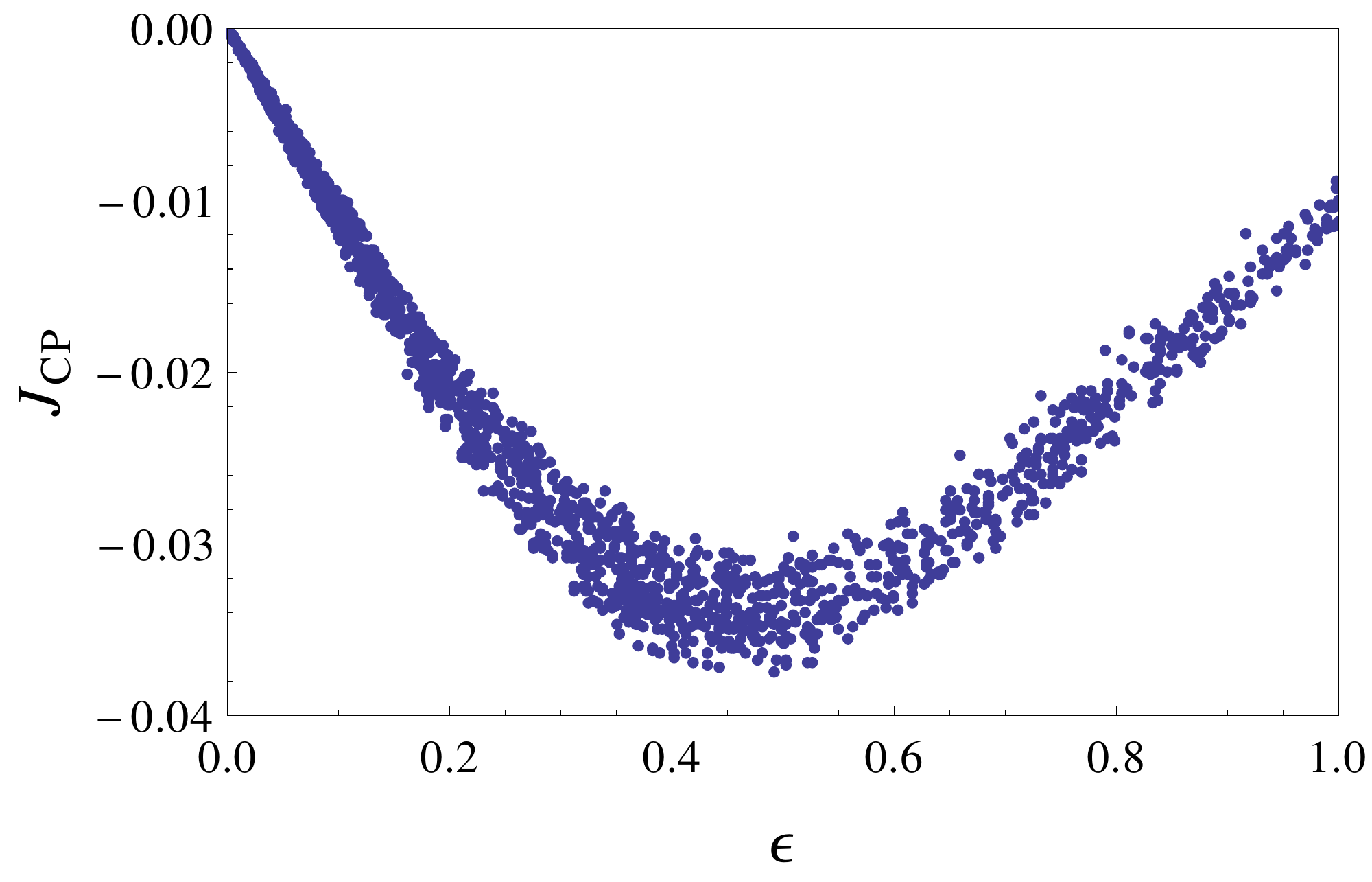}
  \caption{Leptonic CP violation phase $\delta_{CP}$ versus $\epsilon$, the deviation from the reference alignment. We have taken $\alpha = 1.2$. See text. }
  \label{delta}
  \end{figure}

  \noindent Where $\epsilon << 1$. In this limit, one can find a \textbf{correlation between the CP violation parameter, \boldmath $\delta_{CP}$ \unboldmath, and the deviation from the alignment limit which is parametrized by \boldmath $\epsilon$ \unboldmath}. This is shown in the figure \ref{delta}.

\subsection{$A_4$ as flavour symmetry}

This realization of the quarticity model was done in \cite{CentellesChulia:2017koy}. Taking  \boldmath $A_4$\unboldmath \textbf{ as the flavour symmetry} $G$ gives very different results. On the one hand, the model gives some \textbf{strong predictions}
    \begin{eqnarray}
 \theta^\nu_{23} & = & 45^\circ,  \quad \quad \delta^\nu \, = \, \pm \, 90^\circ  \nonumber \\
 \theta^\nu_{12} & = & \rm{arbitrary} \quad \quad  \theta^\nu_{13} \, = \, \rm{arbitrary} 
 \label{unu}
\end{eqnarray}

\noindent While $\theta_{12}^\nu$ and $\theta_{13}^\nu $ are strongly correlated. However, \textbf{this correlation does not allow the fitting of the experimental constraints} for these two angles. The model can be 
\textbf{minimally extended} to include another set of Higgs doublet, which only couples to the up quarks and the neutrinos, while the other Higgs doublet only couples to the d-type quarks and the charged leptons. This way, \textbf{not only 
the mixing parameters can be fitted but also a series of interesting predictions arise}. The charged assignments are \pagebreak

\begin{table}[!h]
\centering
\begin{tabular}{c c c c || c c c c}
  \hline \hline
Fields          \hspace{1cm}    & $SU(2)_L$           \hspace{1cm}   & $A_4$           \hspace{1cm}   &  $Z_4$	    \hspace{1cm}    & Fields          \hspace{1cm}    & $SU(2)_L$           \hspace{1cm}   & $A_4$           \hspace{1cm}   &  $Z_4$	     \\
\hline \hline
$\bar{L}_i$     \hspace{1cm}    & $\mathbf{2}$    \hspace{1cm}    & $\mathbf{3}$    \hspace{1cm}   &  $\mathbf{z}^3$   \hspace{1cm}   &		
$\nu_{e,R}$     \hspace{1cm}    & $\mathbf{1}$    \hspace{1cm}    & $\mathbf{1}$    \hspace{1cm}   &  $\mathbf{z}$      \\
$\bar{N}_{i,L}$ \hspace{1cm}    & $\mathbf{1}$    \hspace{1cm}    & $\mathbf{3}$    \hspace{1cm}   &  $\mathbf{z}^3$   \hspace{1cm}   &        
$\nu_{\mu, R}$  \hspace{1cm}    & $\mathbf{1}$    \hspace{1cm}    & $\mathbf{1'}$   \hspace{1cm}   &  $\mathbf{z}$       \\
$N_{i,R}$       \hspace{1cm}    & $\mathbf{1}$    \hspace{1cm}    & $\mathbf{3}$    \hspace{1cm}   &  $\mathbf{z}$      \hspace{1cm}  &		
$ \nu_{\tau,R}$ \hspace{1cm}    & $\mathbf{1}$    \hspace{1cm}    & $\mathbf{1''}$  \hspace{1cm}   &  $\mathbf{z}$        \\
$ l_{i,R} $     \hspace{1cm}    & $\mathbf{1}$    \hspace{1cm}    & $\mathbf{3}$    \hspace{1cm}   &  $\mathbf{z}$      \hspace{1cm}  &		
$ d_{i,R} $     \hspace{1cm}    & $\mathbf{1}$    \hspace{1cm}    & $\mathbf{3}$    \hspace{1cm}   &  $\mathbf{z}$        \\
$ \bar{Q}_{i,L} $  \hspace{1cm}    & $\mathbf{2}$    \hspace{1cm}    & $\mathbf{3}$    \hspace{1cm}   &  $\mathbf{z}^3$    \hspace{1cm}  &		
$ u_{i,R} $     \hspace{1cm}    & $\mathbf{1}$    \hspace{1cm}    & $\mathbf{3}$    \hspace{1cm}   &  $\mathbf{z}$       	\\
\hline
$\Phi_1^u$    \hspace{1cm}        & $\mathbf{2}$      \hspace{1cm}        & $\mathbf{1}$      \hspace{1cm}  	    &  $\mathbf{1}$        \hspace{1cm} &		
$\chi_i$        \hspace{1cm}        & $\mathbf{1}$      \hspace{1cm}   	    & $\mathbf{3}$    	\hspace{1cm} 	    &  $\mathbf{1}$            \\
$\Phi_2^u$    \hspace{1cm}        & $\mathbf{2}$      \hspace{1cm}        & $\mathbf{1'}$     \hspace{1cm}        &  $\mathbf{1}$        \hspace{1cm}     &		             
$\eta$          \hspace{1cm}        & $\mathbf{1}$	\hspace{1cm}        & $\mathbf{1}$      \hspace{1cm}        &  $\mathbf{z}^2$      	      \\
$\Phi_3^u$    \hspace{1cm}        & $\mathbf{2}$	\hspace{1cm}        & $\mathbf{1''}$    \hspace{1cm}        &  $\mathbf{1}$        \hspace{1cm}     &		
$\zeta$         \hspace{1cm}        & $\mathbf{1}$	\hspace{1cm}        & $\mathbf{1}$      \hspace{1cm}        &  $\mathbf{z}$       	 \\
$\Phi_i^d$      \hspace{1cm}        & $\mathbf{2}$	\hspace{1cm}        & $\mathbf{3}$      \hspace{1cm}        &  $\mathbf{1}$        \hspace{1cm}     &		           
                \hspace{1cm}        & 		 	\hspace{1cm}        & 			\hspace{1cm}        &     	          \\
    \hline
  \end{tabular}
\caption{Charge assignments for leptons, quarks,  scalars 
  ($\Phi_i^u$, $\Phi_i^d$ and $\chi_i$) as well as ``dark matter sector'' 
  ($\zeta$ and $\eta$). Here $\mathbf{z}$ is the fourth root
  of unity, i.e. $\mathbf{z}^4 = 1$.  }
 \label{tab3} 
\end{table}

\begin{figure}[!h]
 \centering
\includegraphics[scale=0.38]{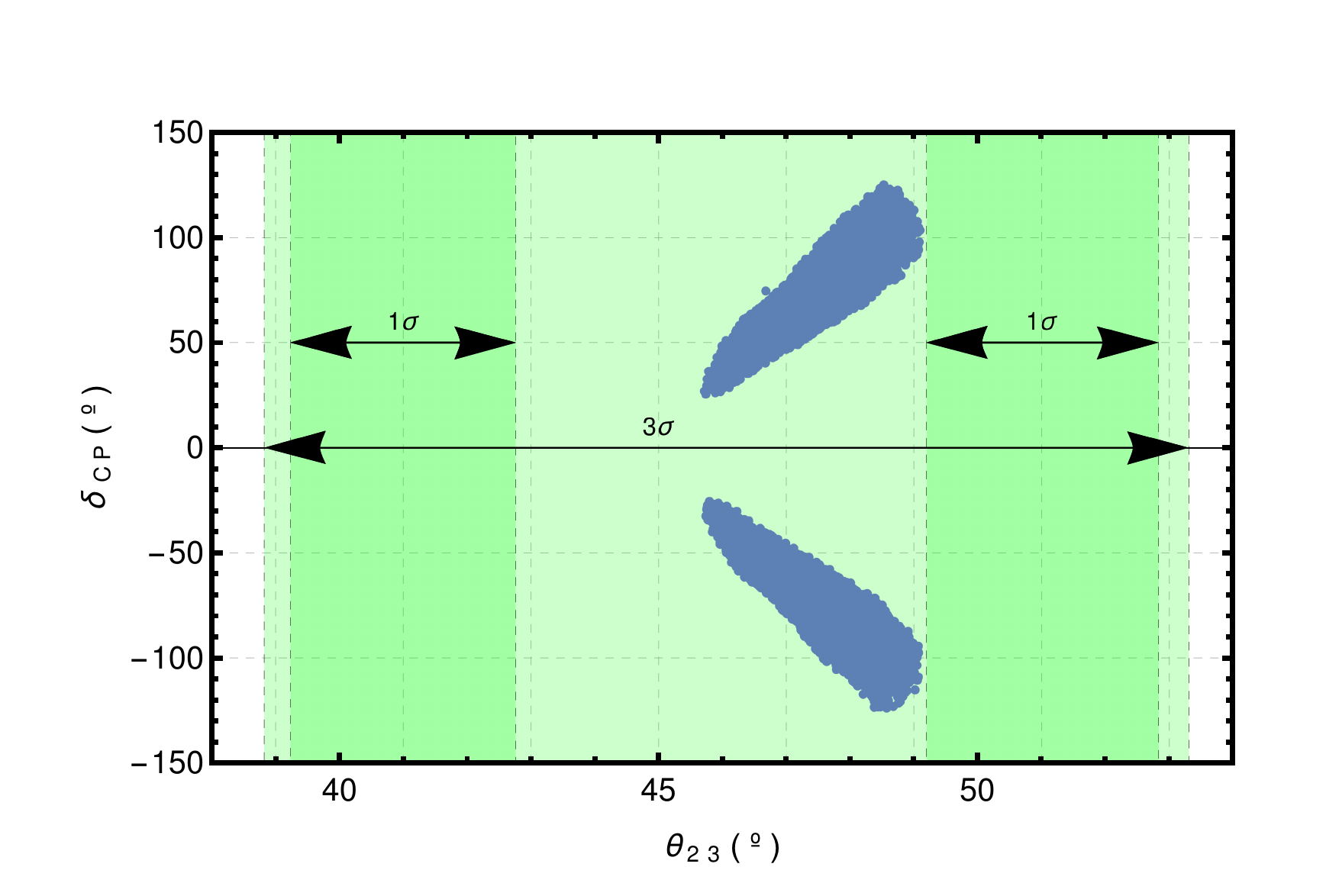}
  \includegraphics[scale=0.36]{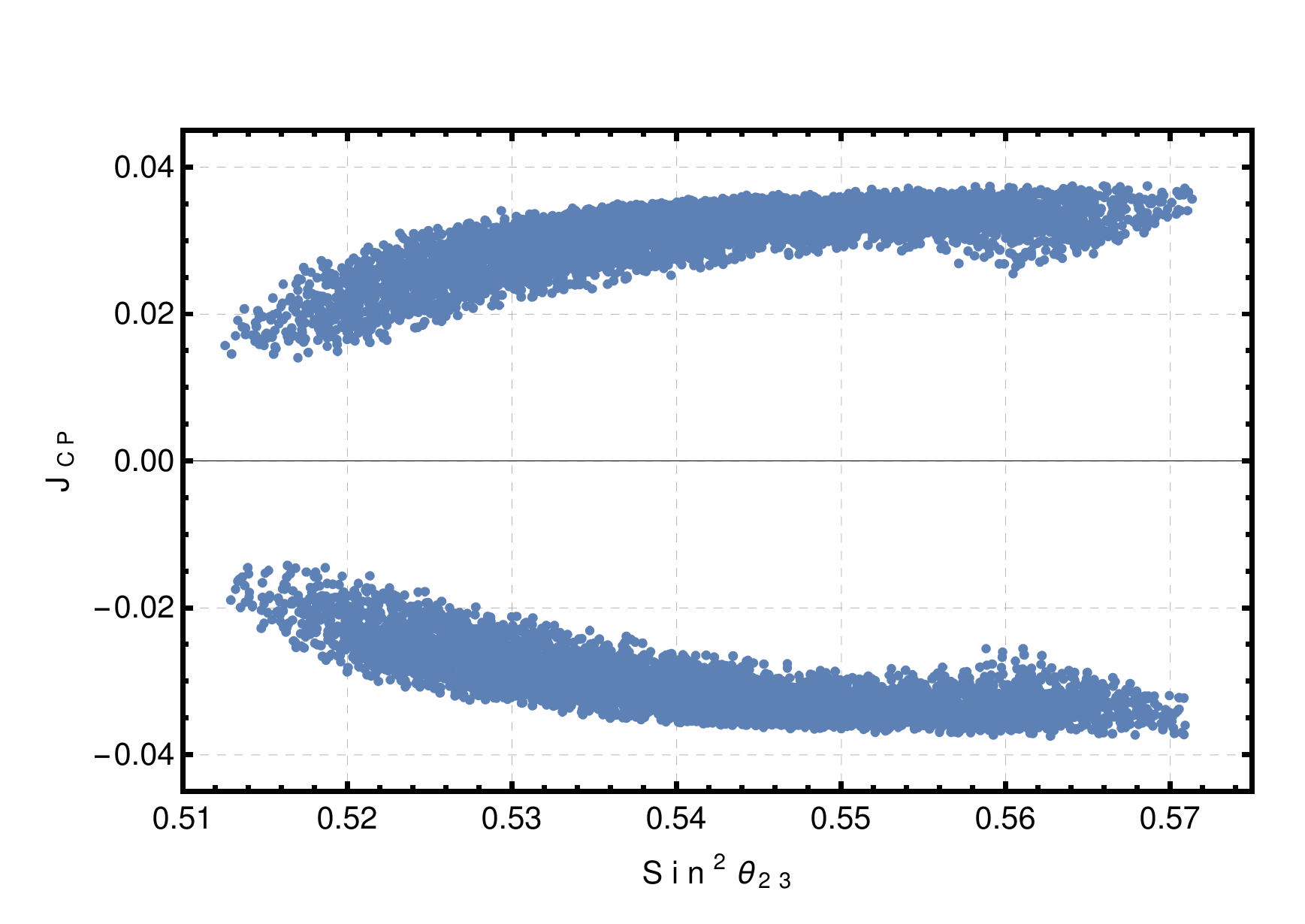}
  \caption{CP violation and $\theta_{23}$ predictions within the
    model. Left panel: $\delta_{CP}$ vs $\theta_{23}$. The green
    regions are the $1\sigma$ (dark) and $3\sigma$ (light) regions for
    $\theta_{23}$ from current oscillation fit. Right panel: Same
    correlation, now showing $J_{CP}$ vs $\sin^2 \theta_{23}$ and
    zooming in the region allowed by the model, fully consistent in
    the $2\sigma$ experimental range.}
    \label{fig:bla}
  \end{figure}

A strong correlation between $\delta_{CP}$ and $\delta_{23}$ appears as shown in figure \ref{fig:bla}
 
  \noindent Also, \textbf{only normal hierarchy of neutrinos is realized}. Moreover, an interesting \textbf{relation between the masses of the quarks and the charged leptons} is also present in the model.
  
  \begin{equation}
  \label{eq:b-tau}
\frac{m_b}{{\sqrt{m_d m_s}}} \approx \frac{m_\tau}{{\sqrt{m_e m_\mu}}}
\end{equation}

\noindent In this case, the flavour symmetry predicts \textbf{(i)} a generalized \textbf{bottom-tau mass relation} involving all families, \textbf{(ii)} small neutrino masses are induced
via a \textbf{type-I seesaw}, \textbf{(iii)} \textbf{CP must be significantly violated} in neutrino oscillations, \textbf{(iv)} the atmospheric angle \boldmath $\theta_{23}$ \unboldmath \textbf{lies in the second octant} 
while showing a \textbf{strong correlation with \boldmath $\delta_{CP}$ \unboldmath}
and \textbf{(v)} only the \textbf{normal neutrino mass ordering is realized.} A more detailed phenomenological analysis of the oscillation predictions of the model has been recently done in \cite{Srivastava:2017sno}.

\section{Conclusions}

We have shown \textbf{a very rich family of predictive models} which can accommodate many different setups while conserving the central features of the model, this is
    
    \begin{itemize}
     \item The quarticity symmetry ${Z}_4$ ensures that neutrinos are Dirac particles and the stability of the DM candidate.
     \item Naturally small neutrino masses are generated via a type-I seesaw.
    \end{itemize}
    
    \noindent Then, the freedom to select the flavour symmetry $G$ leads to a \textbf{whole variety of models with rich predictions}, not only in the neutrino sector but also in the quark and charged lepton sector.

\begin{theacknowledgments}
This article is based on the poster presentation done in the Pontercorvo school 2017 in Prague. The original works were done in collaboration with Jos\'e W. F. Valle, Rahul Srivastava and Ernst Ma.
Work supported by the Spanish grants FPA2017-85216-P and SEV-2014-0398 (MINECO), PROMETEOII/2014/084 (Generalitat Valenciana) and Severo Ochoa IFIC FPI.

\end{theacknowledgments}



\bibliographystyle{aipproc}   


\IfFileExists{\jobname.bbl}{}
 {\typeout{}
  \typeout{******************************************}
  \typeout{** Please run "bibtex \jobname" to obtain}
  \typeout{** the bibliography and then re-run LaTeX}
  \typeout{** twice to fix the references!}
  \typeout{******************************************}
  \typeout{}
 }


\end{document}